# Quantum Process Tomography of a Universal Entangling Gate Implemented with Josephson Phase Qubits

R. C. Bialczak, M. Ansmann, M. Hofheinz, E. Lucero, M. Neeley, A. D. O'Connell, D. Sank, H. Wang, J. Wenner, Matthias Steffen[†], A .N. Cleland, and John M. Martinis[*]

*Department of Physics, University of California, Santa Barbara, CA 93106, USA*

[†]*Present Address: IBM T.J. Watson Research Center, Yorktown, NY 10598, USA.*

[*]*e-mail: martinis@physics.ucsb.edu*

Quantum logic gates must perform properly when operating on their standard input basis states, as well as when operating on complex superpositions of these states. Experiments using superconducting qubits have validated the truth table for particular implementations of *e.g.* the controlled-NOT gate[1,2], but have not fully characterized gate operation for arbitrary superpositions of input states. Here we demonstrate the use of quantum process tomography (QPT)[3,4] to fully characterize the performance of a universal entangling gate between two superconducting quantum bits. Process tomography permits complete gate analysis, but requires precise preparation of arbitrary input states, control over the subsequent qubit interaction, and simultaneous single-shot measurement of the output states. We use QPT to measure the fidelity of the entangling gate and to quantify the decoherence mechanisms affecting the gate performance. In addition to demonstrating a promising fidelity, our entangling gate has a on/off ratio of 300, a level of adjustable coupling that will become a requirement for future high-fidelity devices. This is the first solid-state demonstration of QPT in a two-qubit system, as solid-state process tomography has previously only been demonstrated with single qubits[5,6].

Universal quantum gates are the key elements in a quantum computer, as they provide the fundamental building blocks for encoding complex algorithms and operations. Single qubit rotations together with the two-qubit controlled-NOT (CNOT) are known to provide a universal set of gates[7]. Here, we present the complete characterization of a universal entangling gate, the square root of *i*-SWAP (SQ*i*SW)[8], from which gates such as the CNOT can be constructed[9]. The SQ*i*SW is a "natural" two-qubit gate as it directly results from capacitive coupling of superconducting qubits, yielding qubit coupling of the general form $\sigma_{Ax}\sigma_{Bx}$ or $\sigma_{Ay}\sigma_{By}$, where $\sigma_{x,y}$ are the Pauli spin operators for qubits A and B[10,11]. Under the rotating wave approximation, the corresponding interaction Hamiltonian has the form $H_{int} = \hbar(g/2)(|01\rangle\langle 10| + |10\rangle\langle 01|)$, where $|01\rangle = |0\rangle_A \otimes |1\rangle_B$ and *g* is the coupling strength that depends on design parameters.

When the two qubits are placed on-resonance the two-qubit states are coupled by $H_{int}$ as shown in Fig. 1a. The amplitudes of these states then oscillate in time, as described (in the rotating frame) by the unitary transformation

$$U_{int} = \begin{bmatrix} 1 & 0 & 0 & 0 \\ 0 & \cos(gt/2) & -i\sin(gt/2) & 0 \\ 0 & -i\sin(gt/2) & \cos(gt/2) & 0 \\ 0 & 0 & 0 & 1 \end{bmatrix},$$

where *t* is the interaction time, and the representation is in the two-qubit basis set $\{|00\rangle, |01\rangle, |10\rangle, |11\rangle\}$. For an interaction time $gt=\pi$, the state amplitudes are swapped, such that $|01\rangle \rightarrow -i|10\rangle$ and $|10\rangle \rightarrow -i|01\rangle$. The SQ*i*SW gate is formed by coupling for one-half this time, $gt=\pi/2$, producing cosine and sine matrix elements with equal magnitudes, thus entangling the qubits. When the qubits are off resonance by an energy $|\Delta| \gg g$ (Fig 1b), the off-diagonal elements in $U_{int}$ are small and have average amplitude $g/\Delta$, effectively turning off the qubit-qubit interaction.

The electrical circuit for the capacitively-coupled Josephson phase qubits[12,13] used in this experiment is shown in Fig. 1c. Each phase qubit is a non-linear resonator built from a Josephson inductance and an external shunting capacitance. When biased close to the critical current, the junction and its parallel loop inductance *L* give rise to a cubic potential whose energy eigenstates are unequally spaced. The two lowest levels are used as the qubit states $|0\rangle$ and $|1\rangle$, with transition frequency $\omega_{10}$. This frequency can be adjusted independently for each qubit via the bias current $I_{bias}^{A,B}$. Each qubit's state is detected via a single-shot measurement[14,15], using a fast pulse $I_Z^{A,B}$ combined with read-out using an on-chip SQUID detector.

State preparation and tomography use single-qubit logic operations, corresponding to rotations about the *x*, *y*, and *z*-axes of the Bloch sphere[15]. Rotations about the *z*-axis are produced by fast (~ ns) current pulses $I_Z^{A,B}(t)$, which adiabatically change the qubit frequency, turning on and off the interaction and leading to phase accumulation between the $|0\rangle$ and $|1\rangle$ states. Rotations about any axis in the *x-y* plane are produced by microwave pulses resonant with each qubit's transition frequency, applied via $I_{\mu w}^{A,B}(t)$. The phase of the microwave pulses defines the rotation axis in the *x-y* plane, and the pulse duration and amplitude control the rotation angle. In previous work[16], such single-qubit gates were shown to have fidelities of 98%, limited by the energy relaxation $T_1$ and dephasing $T_2$ times, which for this device were measured to be 400 ns and 120 ns, respectively.

The experimental design was chosen to give qubit frequencies $\omega_{10}^{A,B}/2\pi \cong 5.5$ GHz. The strength of the coupling $g = (C_c/C)\omega_{10}^{A,B}$ was set by the coupling and qubit capacitances $C_c \approx 2$ fF and $C \approx 1$ pF, respectively. The coupling interaction is turned on and off by changing the relative qubit frequency $\Delta = \omega_{10}^A - \omega_{10}^B$ through an adjustment of the qubit B bias $I_{bias}^B$. A large detuning of $\Delta_{off}/2\pi \approx 200$ MHz was used to turn off the gate, yielding a small amplitude in the off-diagonal coupling $g/\Delta_{off} \approx 0.055$.

We first characterize the coupling by measuring the time dynamics of the entangling swap operation $|01\rangle \leftrightarrow |10\rangle$ as depicted in Fig. 2a. Initially, both qubits are tuned off-resonance by 200 MHz and allowed to relax to the $|00\rangle$ state. A $\pi$ pulse on qubit A then produces the $|10\rangle$ state. A current pulse $I_Z^B(t)$ applied to $I_{bias}^B$ brings the qubits within a frequency $\Delta$ of resonance. After an interaction time $t_f$, the bias $I_Z^B$ is reset to the original 200 MHz detuning and both qubits are then measured. Averaging over 1200 events gives the probabilities for the four possible final states $|00\rangle$, $|01\rangle$, $|10\rangle$ and $|11\rangle$. The swapping behaviour for the states $|01\rangle$ and $|10\rangle$ as a function of $t_f$ is displayed in Fig. 2b. On resonance ($\Delta = 0$), the swapping frequency between $|01\rangle$ and $|10\rangle$ gives an accurate measurement of the coupling strength $g/2\pi = 11$ MHz.

The amplitude of the swapping oscillations decreases with detuning as expected. In Fig. 2c we plot the peak-to-peak change in swap probability as a function of detuning $\Delta$, compared to the theoretical prediction. Apart from a small reduction in the amplitude arising from imperfect measurement fidelity, the data is in good agreement with theory. At detunings $|\Delta|/2\pi > 50$ MHz, the swap amplitude is small and cannot be distinguished from the noise floor. For a detuning bias of $\Delta/2\pi = 200$ MHz, we compute the probability ratio $(\Delta/g)^2 = (200/11)^2 = 300$, a figure of merit for the on/off coupling ratio.

We fully characterize the SQ$i$SW gate using quantum process tomography (QPT)[3,4]. This involves preparing the qubits in a spanning set of input basis states, operating with the gate on this set of states, and then performing complete state tomography on the output. As illustrated in Fig. 3a, we first perform quantum state tomography[15,17] on the input state $|01\rangle$, which involves measuring the state along the $x$, $y$ and $z$ Bloch sphere axes of each qubit, in nine separate experiments. We then operate on the $|01\rangle$ input state with SQ$i$SW, and perform complete state tomography on the output. These measurements allow for the evaluation of the two-qubit density matrix. This entire process is repeated 16 times in total, using four distinct input states for each qubit, chosen from the set $\{|0\rangle, |1\rangle, |0\rangle+|1\rangle, |0\rangle+i|1\rangle\}$. In Fig. 3b, we display the density matrix resulting from this tomography for the input state $(|0\rangle+i|1\rangle)\otimes(|0\rangle+i|1\rangle)$. From this complete set of measurements, we reconstruct the 16 by 16 $\chi$ matrix, whose indices correspond to the Kronecker product of the operators $\{I, \sigma_x, -i\sigma_y, \sigma_z\}$ for each qubit[3].

In a QPT experiment[18,19] errors arise from the entangling gate and errors in measurement. Since we are interested in the quality of the entangling gate itself, we have calibrated out errors due to measurement[14]. As described in the Supplementary Information, measurement errors arise from both a misidentification of the $|0\rangle$ and $|1\rangle$ states, and measurement crosstalk, where a measurement of $|1\rangle$ in one qubit increases the probability of a $|1\rangle$ measurement in the second qubit[14]. By performing additional calibration experiments, we are able to determine the probabilities for these errors and correct the probabilities of the $|00\rangle$, $|01\rangle$, $|10\rangle$ and $|11\rangle$ final states.

In Fig. 4 we show the corrected $\chi$ matrix for our SQ$i$SW gate. In both the real and imaginary parts of the $\chi$ matrix, we observe non-zero matrix elements in locations where such elements are expected, in qualitative agreement with theory. Quantitative comparison is obtained by calculating the process fidelity, $0 < F_p < 1$, which gives a measure of how close the measured $\chi$ matrix is to theoretical expectations[20]. For the SQ*i*SW gate demonstrated here, with measurement calibration taken into account, we find $F_p^c = Tr(\chi_t \chi_e) = 0.65$, where $\chi_t$ and $\chi_e$ are the theoretical and experimental $\chi$ matrices, respectively. Without calibration, we compute $F_p = 0.51$; the uncorrected $\chi$ matrix is shown in the Supplementary Information.

Errors in our SQ*i*SW gate primarily arise because the time for the experiment (~50 ns) is not significantly shorter than the $T_2$ dephasing time of 120 ns. This is confirmed using a recent theory[21] by Kofman *et al.*, which includes the effects of dephasing and decoherence on the SQ*i*SW $\chi$ matrix. In particular, the elements marked with "*" and "o" in Fig. 4 are non-zero due to energy relaxation and dephasing, respectively. Using this theory we can estimate our single qubit dephasing time $T_2 = (3\pi + 2)/16g\chi_e^{IZ,IZ}$. With the measured real part $\chi_e^{IZ,IZ} = 0.09$, we find $T_2 = 110$ ns, in close agreement with the value mentioned above obtained from Ramsey experiments. We also estimate the degree of correlation of the dephasing noise between the coupled qubits using $\kappa \approx \chi_e^{IZ,ZI}/\chi_e^{IZ,IZ} - [(\pi-2)/(3\pi+2)]$. Our measurement of $\chi_e^{IZ,ZI} \approx 0.02$ yields $\kappa \cong 0.1$, indicating that the dephasing is mostly uncorrelated. This is in agreement with previous work[22,23] that found a dephasing mechanism local to the individual qubits.

In conclusion, we have demonstrated the universal SQ*i*SW gate, from which the CNOT and other more complex gates can be constructed. Using quantum process tomography, we obtained a process fidelity of 65% for this gate. By analyzing the $\chi$ matrix obtained via QPT, we confirmed dephasing times obtained via separate, single-qubit Ramsey experiments and found that the correlation of dephasing between qubits is small. Finally, we showed that our implementation of the SQ*i*SW gate has a high on/off ratio of 300, which is needed for high-fidelity entangling gates.

# Acknowledgements

Devices were made at the UCSB Nanofabrication Facility, a part of the NSF-funded National Nanotechnology Infrastructure Network. We thank A.N. Korotkov for discussions.

This work was supported by IARPA (grant W911NF-04-1-0204) and by the NSF (grant CCF-0507227).

**Figure 1** Energy-level diagram with coupling interaction turned on (a) and off (b). (a) When qubits are on-resonance ($\Delta = 0$), their interaction swaps the populations of the $|01\rangle$ and $|10\rangle$ states at a frequency given by the coupling strength $g$. The Bloch sphere representation of the $|01\rangle$ and $|10\rangle$ subspace shows state rotation (dashed line) about the *x*-axis due to the interaction $g$. (b) When off-resonance $|\Delta| \gg g$, qubit swapping (dashed line) is effectively turned off. (c) Electrical schematic for capacitively-coupled phase qubits. Each qubit junction (single cross) with critical current $I_0$ is shunted by an external capacitor $C$ and inductor $L$. An interdigitated capacitor $C_c \sim 2$ fF couples the qubits, yielding an interaction $g/2\pi = 11$ MHz. Qubit bias is through $I_{\text{bias}}$, microwave control through $I_{\mu w}$, and qubit readout is performed using a three-junction SQUID (three crosses) read out by $V_{\text{SQ}}$. Fast pulses on the $I_z$ lines bias the qubits on- and off-resonance.

**Figure 2** Characterization of coupling interaction and measurement of on/off ratio. (a) Sequence of operations: The $|10\rangle$ state is prepared by applying a 16 ns $\pi$ pulse to qubit A, and immediately followed by a fast pulse $I_Z^B$ that places qubit B close to resonance with qubit A (detuning $\Delta$). After the qubits interact for a time $t_f$, the state occupation probabilities for each qubit are measured simultaneously. (b) Measured occupation probabilities P$_{10}$ and P$_{01}$ versus detuning $\Delta$ and interaction time $t_f$. The oscillation period at $\Delta = 0$ yields the measured coupling $g/2\pi = 11$ MHz. The amplitude of oscillations decreases with detuning as expected. (c) Peak-to-peak swapping amplitude versus detuning $\Delta$, plotted with the predicted dependence $g^2/(g^2 + \Delta^2)$. The vertical scale of the latter is adjusted to match the on-resonance amplitude at $\Delta = 0$. Determination of the swapping probability is limited to $> 6 \cdot 10^{-2}$ by measurement noise. The calculated on/off ratio is indicated by the vertical arrow.

**Figure 3.** Quantum state tomography for two sets of input and output states. Control sequences are shown in left panels, where the SQ*i*SW gate is represented by crosses connecting both qubits. Qubit state measurements along the *z*, *y* and *x* axes are performed by no rotation (*I*) or π/2 rotations about *x* and *y*, respectively, followed by measuring the probability of the qubit $|1\rangle$ state. Right panels show the real and imaginary parts of the density matrices obtained in this way. The experimental data (with no corrections) and theory (with no decoherence) are shown as solid and transparent bars, respectively. Panels (a) and (b) are for input states $|01\rangle$ and $(|0\rangle + i|1\rangle) \otimes (|0\rangle + i|1\rangle)$, generated by a π pulse about *x* applied to qubit B, and by π/2 pulses about *x* applied to both qubits A and B, respectively.

**Figure 4** Real and imaginary parts of the reconstructed $\chi$ matrix for the SQ*i*SW gate, obtained from 16 possible input states $\{|0\rangle, |1\rangle, |0\rangle + |1\rangle, |0\rangle + i|1\rangle\} \otimes \{|0\rangle, |1\rangle, |0\rangle + |1\rangle, |0\rangle + i|1\rangle\}$. Experimental data are shown as solid bars. Transparent bars give the theoretically expected $\chi$ matrix, which does not include effects due to decoherence. Calibrations from measurement were accounted for in this analysis (see Supplementary Information). The matrix elements of $\chi$ that are non-zero due to energy relaxation and dephasing are marked with a "*" and "o" symbol, respectively.

# Figure 1

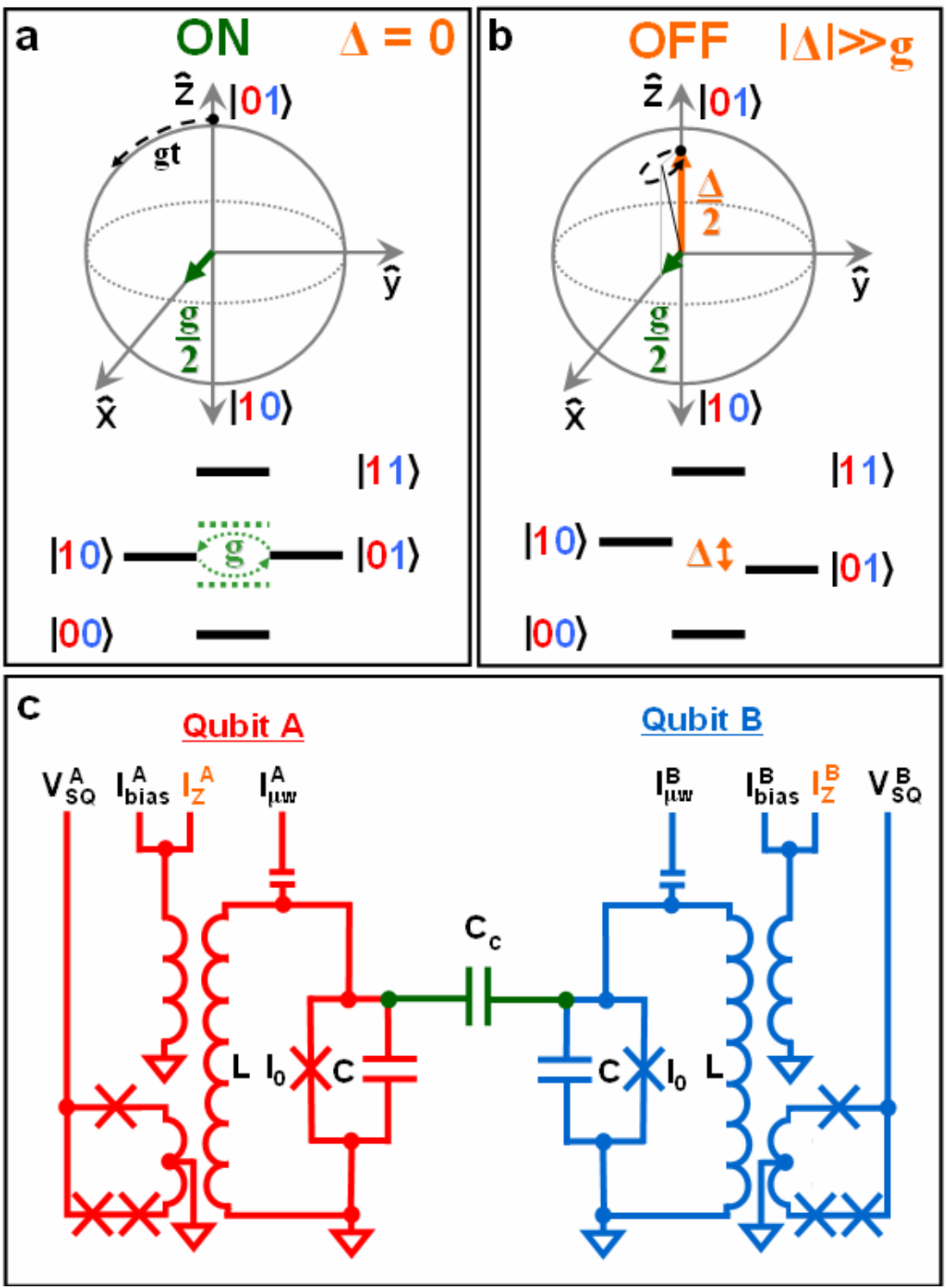

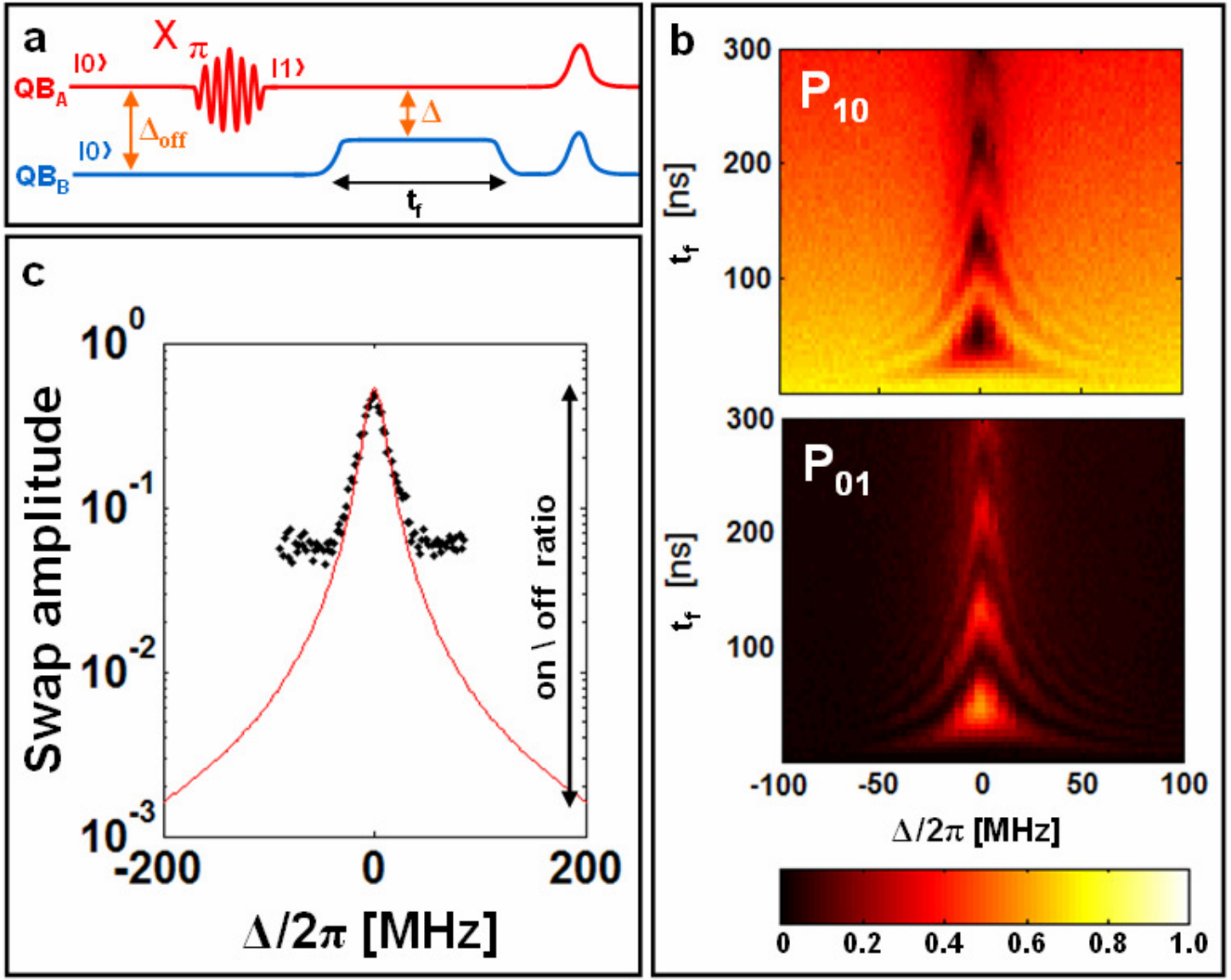

**Figure 3**

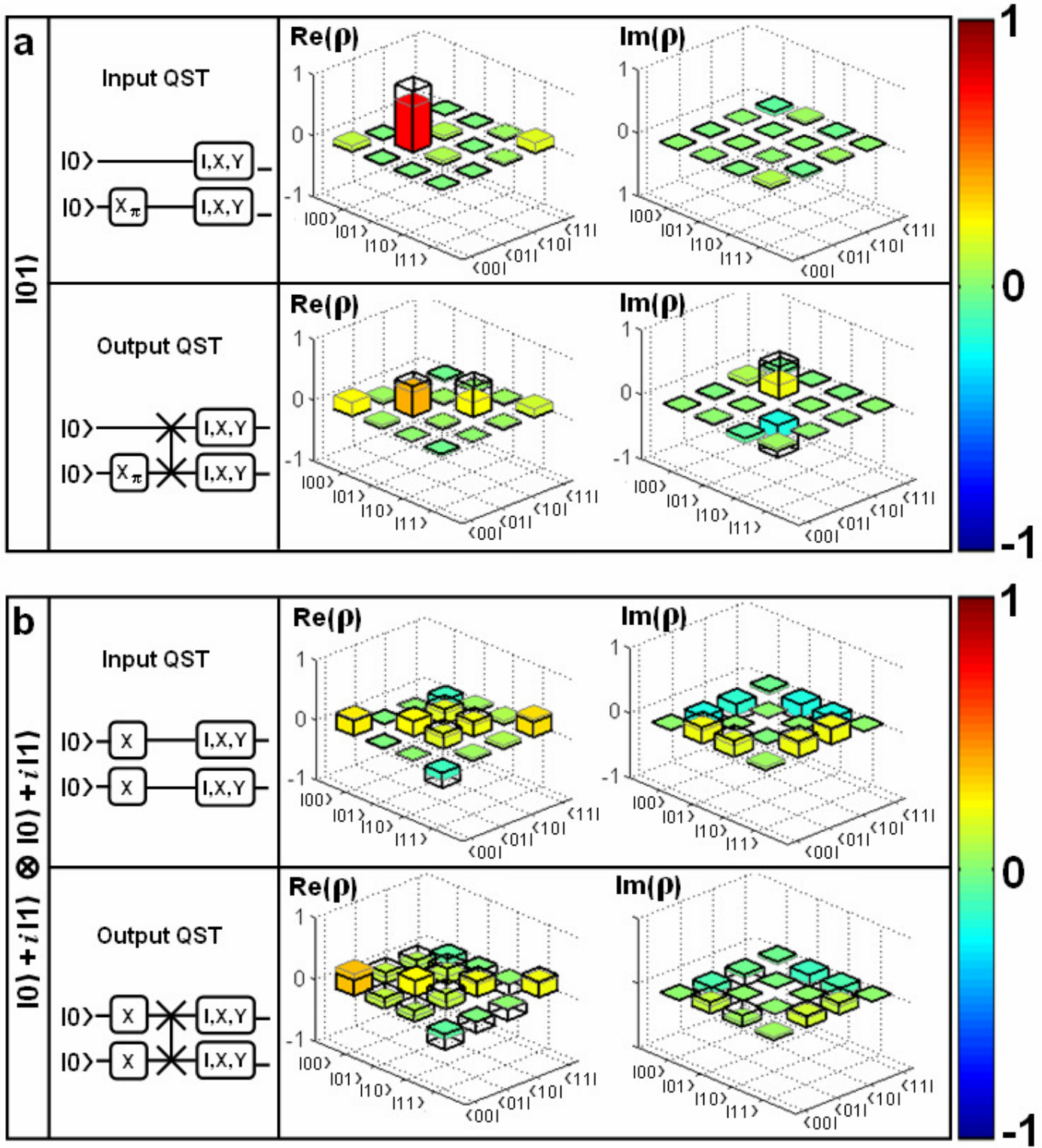

**Figure 4**

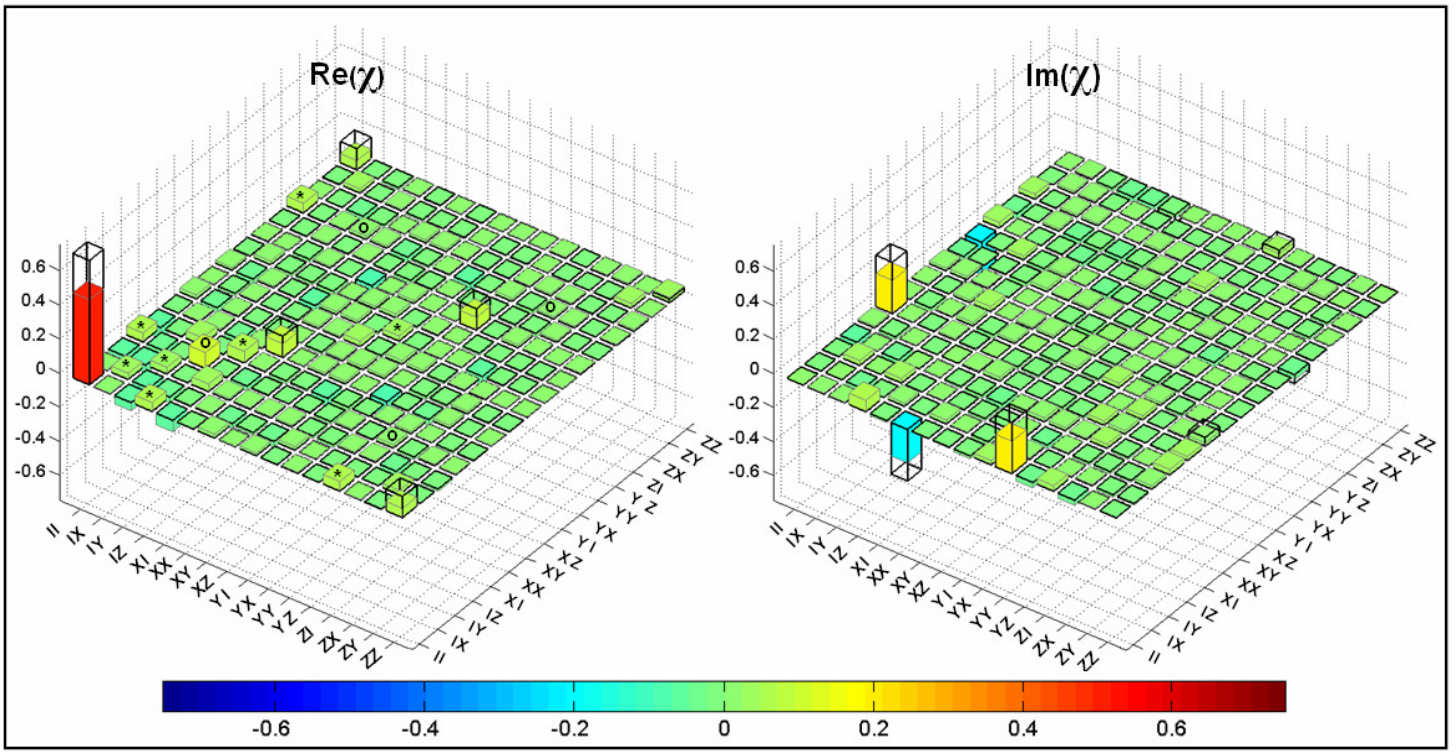

## Supplementary Information

**Calibration of Measurement Errors**

The measurement errors in this experiment primarily arise from two sources. Because their origin is understood and the errors vary in a predictable way with parameters and biasing, the errors can be reliably removed using calibration procedures.

The two dominant error mechanisms are measurement crosstalk and measurement fidelity. Defining the measurement probabilities $P_{AB}$ of qubits A and B with the column vector $(P_{00}, P_{01}, P_{10}, P_{11})^T$, the intrinsic (actual) probabilities $P_i$ will give measured probabilities $P_m$ according to the matrix equation $P_m = XFP_i$, where $X$ is the correction matrix due to measurement crosstalk and $F$ is the correction due to measurement fidelity. The order of the matrices reflects the fact that errors in fidelity generate crosstalk (see below). By measuring the correction matrices, the intrinsic probabilities can be calculated from the measured values by the inverted relation $P_i = F^{-1} X^{-1} P_m$.

The procedure for calibrating measurement fidelity for single qubits has been discussed previously[15]. The measurement process, which depends on the $|0\rangle$ qubit state not tunneling and the $|1\rangle$ state tunneling, has small errors from $|0\rangle$ tunneling and $|1\rangle$ not tunneling. Defining $f_0$ and $f_1$ as the probabilities to correctly identify the state in $|0\rangle$ and $|1\rangle$, respectively, the measurement fidelity matrix for two qubits is given by

$$F = \begin{pmatrix} f_0 & 1-f_1 \\ 1-f_0 & f_1 \end{pmatrix}_A \otimes \begin{pmatrix} f_0 & 1-f_1 \\ 1-f_0 & f_1 \end{pmatrix}_B$$

$$= \begin{pmatrix} f_{0A}f_{0B} & f_{0A}(1-f_{1B}) & (1-f_{1A})f_{0B} & (1-f_{1A})(1-f_{1B}) \\ f_{0A}(1-f_{0B}) & f_{0A}f_{1B} & (1-f_{1A})(1-f_{0B}) & (1-f_{1A})f_{1B} \\ (1-f_{0A})f_{0B} & (1-f_{0A})(1-f_{1B}) & f_{1A}f_{0B} & f_{1A}(1-f_{1B}) \\ (1-f_{0A})(1-f_{0B}) & (1-f_{0A})f_{1B} & f_{1A}(1-f_{0B}) & f_{1A}f_{1B} \end{pmatrix},$$

where $\otimes$ is the tensor product. We measure these fidelities by biasing only one qubit into operation, and then measuring the tunneling probabilities for the $|0\rangle$ state and the $|1\rangle$ state, with the latter produced by a microwave π-pulse optimized for the largest tunneling probability. This calibration depends on accurately producing the $|1\rangle$ state, which we have demonstrated can be done with 98% accuracy. The 2% error arises from $T_1$ energy decay, which can be measured and corrected for in the calibration.

Measurement crosstalk for two capacitively coupled Josephson phase qubits has been studied and understood in previous work[14]. The crosstalk mechanism arises from the release of energy when one qubit tunnels, which then excites the second qubit and increases its probability to tunnel. The amount of crosstalk typically increases with time delay, so that optimal performance occurs when the two qubits are measured simultaneously. For this mechanism, crosstalk contributes when one qubit state is measured as 1, causing the other qubit state, when in the 0 state, to have probability $x$ to be excited and thus measured in the 1 state. The matrix describing measurement crosstalk for both qubits is thus

$$X = \begin{pmatrix} 1 & 0 & 0 & 0 \\ 0 & 1-x_{BA} & 0 & 0 \\ 0 & 0 & 1-x_{AB} & 0 \\ 0 & x_{BA} & x_{AB} & 1 \end{pmatrix},$$

where $x_{AB}$ ($x_{BA}$) is the probability of the $|1\rangle$ state of qubit A (qubit B) exciting a $|0\rangle \rightarrow |1\rangle$ transition on qubit B (qubit A).

The two unknowns in the $X$ matrix can be directly determined from the 3 independent equations in $P_m = X(FP_i)$, where $FP_i$ is obtained from the $F$ matrix calibration procedure described above.

A more robust method is to compare the differences in tunneling of the first qubit caused by a change in tunneling of the second. From the four measurement probabilities $P_{00}$, $P_{01}$, $P_{10}$, and $P_{11}$, we extract for each qubit independent probabilities to be in the $|1\rangle$ state by "tracing out" the other qubit

$P_{1A} \equiv P_{10} + P_{11}$

$P_{1B} \equiv P_{01} + P_{11}$

We measure $P_{1A}(00)$ and $P_{1B}(01)$ for the two cases where we prepare the initial states $|00\rangle$ and $|01\rangle$, respectively. Using the correction matrices for $X$ and $F$, we calculate

$$\frac{P_{1A}(01) - P_{1A}(00)}{P_{1B}(01) - P_{1B}(00)} = \frac{f_{0A}}{1-(1-f_{0A})x_{AB}} \cdot x_{BA} ,$$
$$\cong f_{0A} x_{BA}$$

where the approximate result arises from both correction terms in the denominator, $1-f_{0A}$ and $x_{BA}$, being small. This result holds even if the states $|00\rangle$ and $|01\rangle$ are not prepared perfectly, as we calculate the ratio of the change in probabilities. A similar result for $f_{0B}x_{AB}$ is obtained for the initial states $|00\rangle$ and $|10\rangle$.

We also perform a consistency check on the measurements of $x_{AB}$ and $x_{BA}$ for the simple case of measuring only the $|00\rangle$ state when $f_{0A}, f_{0B} \neq 0$. Here, a general solution is not possible as there are four unknowns $f_{0A}$, $f_{0B}$, $x_{AB}$, and $x_{BA}$ and only three equations for the probabilities. However, by assuming a fixed ratio between the two crosstalk parameters $k = x_{BA}/x_{AB}$, a solution can be found:

$$x_{AB} = \frac{P_{00} + kP_{00} - k + kP_{10} + P_{01} - 1 - \sqrt{((1-P_{00}-P_{01}) - k(1-P_{00}-P_{10}))^2 + 4kP_{10}P_{01}/P_{00}}}{2k(P_{00}-1)}$$

For the device measured here, we found measurement fidelities that were near unity: $f_{0A} = 0.95$ and $f_{1A} = 0.95$ for qubit A, and $f_{0B} = 0.93$ and $f_{1B} = 0.93$ for qubit B. Crosstalk was measured to be $x_{AB} = x_{BA} = 0.117$.

**Uncalibrated $\chi$ Matrix For The SQiSW Gate**

Below we give the $\chi$ matrix for the SQ*i*SW gate without calibrating out the effects of measurement visibility and crosstalk. The uncalibrated process fidelity is $F_p = Tr(\chi_{theory} \chi_{exp}) = 0.51$:

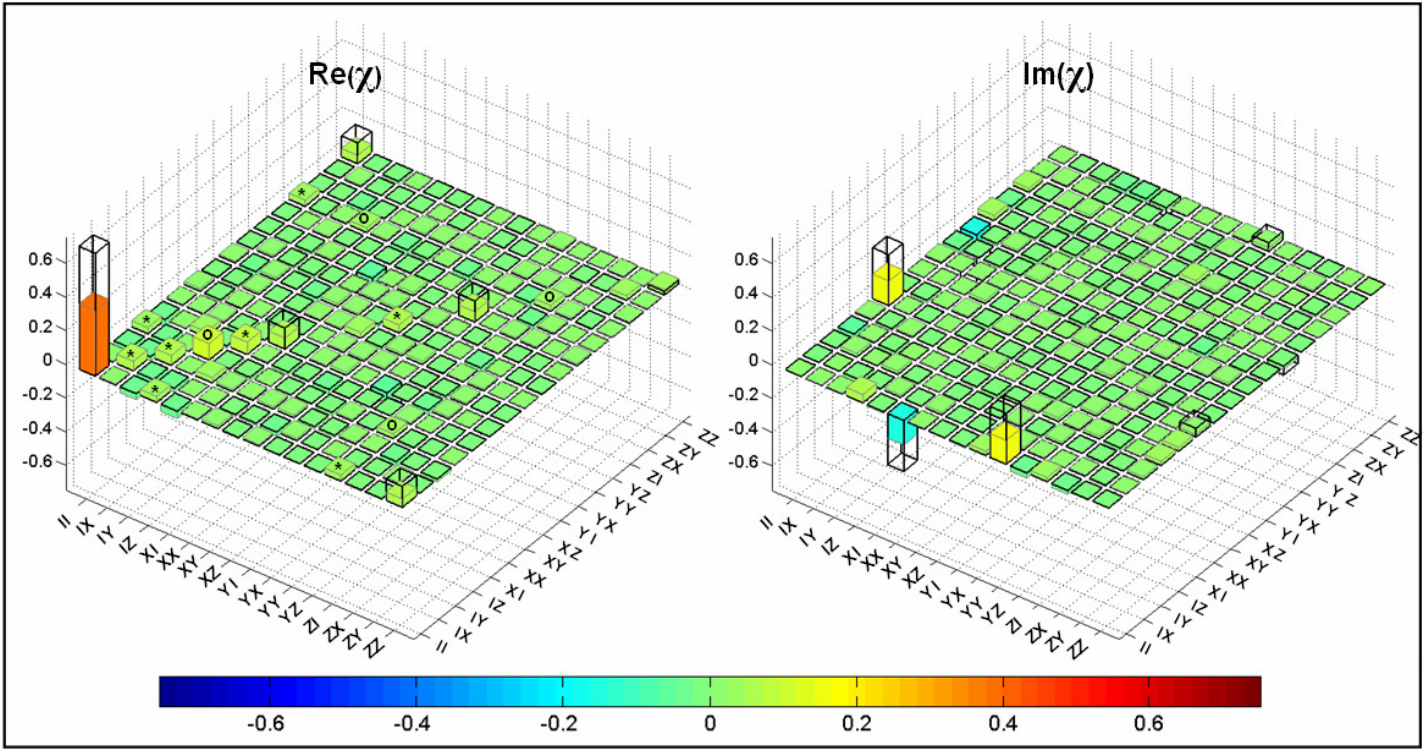